# Disconnection-Mediated Twin/Twin-Junction Migration in FCC metals

Mingjie Xu[1,2,3+], Kongtao Chen[4+], Fan Cao[1], Leonardo Velasco Estrada[5], Thomas M. Kaufman[1], Fan Ye[1], Horst Hahn[5], Jian Han[6*], David J. Srolovitz[7,8*], Xiaoqing Pan[1,2,3*]

[1] Department of Materials Science and Engineering, University of California, Irvine, Irvine, CA 92697
[2] Department of Physics and Astronomy, University of California, Irvine, Irvine, CA 92697
[3] Irvine Materials Research Institute (IMRI), University of California, Irvine, Irvine, CA 92697
[4] Department of Materials Science and Engineering, University of Pennsylvania, Philadelphia, PA 19104
[5] Institute of Nanotechnology, Karlsruhe Institute of Technology (KIT), Karlsruhe, Germany
[6] Department of Materials Science and Engineering, City University of Hong Kong, Kowloon, Hong Kong, China
[7] Department of Mechanical Engineering, The University of Hong Kong, Pokfulam, Hong Kong SAR, China
[8] International Digital Economy Academy (IDEA), Shenzhen, China.

[+]Equal contributions

[*]Corresponding authors: Jian Han (jianhan@cityu.edu.hk) David J. Srolovitz (srol@hku.hk) Xiaoqing Pan (xiaoqinp@uci.edu)

**Author Contributions:** X.P. designed the experiments. X.P. and D.J.S. directed the project. L.V.E. synthesized the samples. T.M.K., F.C., M.X. and F.Y. performed in situ electron microscopy experiments. K.C. performed the simulations. K.C. and J.H. developed theory. T.M.K., F.C., and M.X. participated in experimental data analysis. T.M.K and K.C. wrote the manuscript and prepared figures. J.H., H.H., D.J.S, and X.P. revised the manuscript. All authors participated in discussions and know the implications of the work.

**Competing Interest Statement:** The authors declare no competing interests.

**Classification:** PHYSICAL SCIENCES/Engineering

**Abstract**

We present the results of novel, time-resolved, *in situ* HRTEM observations, molecular dynamics (MD) simulations, and disconnection theory that elucidate the mechanism by which the motion of grain boundaries (GBs) in polycrystalline materials are coupled through disconnection motion/reactions at/adjacent to GB triple junctions (TJs). We focus on TJs composed of a pair of coherent twin boundaries (CTBs) and a Σ9 GB. As for all GBs, disconnection theory implies that multiple modes/local mechanisms for CTB migration are possible and that the mode selection is affected by the nature of the driving force for migration. While we observe (HRTEM and MD) CTB migration through the motion of pure steps driven by chemical potential jump, other experimental observations (and our simulations) show that stress-driven CTB migration occurs through the motion of disconnections with a non-zero Burgers vector; these are pure-step and twinning-partial CTB migration mechanisms. Our experimental observations and simulations demonstrate that the motion of a GB drags its delimiting TJ and may force the motion of the other GBs meeting at the TJ. Our experiments and simulations focus on two types of TJs composed of a pair of CTBs and a Σ9 GB; a 107° TJ readily migrates while a 70° TJ is immobile (experiment, simulation) in agreement with our disconnection theory even though the intrinsic mobilities of the constituent GBs do not depend on TJ-type. We also demonstrate that disconnections may be formed at TJs (chemical potential jump/stress driven) and at GB/free surface junctions (stress-driven).

**Significance Statement**

Grain boundary (GB) kinetics is important to plasticity of polycrystalline metals. GB kinetics in polycrystalline materials is coupled through disconnection motion at GB triple junctions (TJs). We study the migration mechanisms of the most common TJs, 70° and 107° coherent twin junctions (composed of a pair of coherent twin boundaries (CTBs) and a Σ9 GB), by HRTEM observations, molecular dynamics (MD) simulations, and disconnection theory. We find that their migration mechanism depends on the nature of the driving force: CTBs migrate by pure steps nucleated near TJ under chemical potential jump, and by partial dislocations with finite Burgers vector nucleated at TJ or GB/free surface junctions under shear stress. The 107° TJ readily migrates while a 70° TJ is immobile.

**Introduction**

Most materials are polycrystalline and their properties are greatly influenced by the network of grain boundaries (GBs). An accurate understanding of GB kinetics is necessary for the control of processing-structure-property relationships to design and model optimized materials[1]. Additionally, the plastic deformation of polycrystalline materials is also strongly influenced by processes occurring at GBs, including GB sliding, GB migration, as well as direct interaction with dislocations[2,3]. In order to accurately model GB kinetics and properties, we must focus on the underlying mechanisms of GB dynamics; i.e., the motion of disconnections within GBs. Disconnections are line defects constrained to GBs that have both dislocation and step characters[4–10]. The motion of a GB occurs by the nucleation and migration of disconnections that are controlled by factors associated with either their step or dislocation character that couple these defects to driving forces that include chemical potential jumps (e.g. curvature, temperature gradient, defect concentration gradient) and mechanical loading[5]. Triple junctions (TJs), are the one-dimensional intersections of three GB planes; they provide boundary conditions for GB migration and can be locations where disconnections pile-up or form (elastic fields from disconnection pileups may restrict TJ and GB motion)[7]. Such disconnection-mediated GB/TJ migration phenomena have been observed directly by *in situ* conventional transmission electron microscopy (TEM)[11–13], and more recently at atomic resolution in high resolution transmission electron microscopy (HRTEM) experiments[14,15].

Twinning is an important deformation mechanism in metals and alloys[16–27], especially for coarse-grained face-centered-cubic (FCC) metals with low stacking fault energies (e.g., Ag[28]) and for nanocrystalline FCC metals[29]. In coarse-grained metals, twins provide a deformation mode that is complementary to dislocation plasticity. Twinning in FCC metals is commonly associated with the glide of identical partial dislocations on successive (111) plane resulting in a net strain. In nanocrystalline metals, twins act to stabilize the microstructure and their motion during grain growth produces no net (macroscopic) strain via random (RAP) or cooperative activation of partials[30–32]. The growth of twinned domains by the advancement of Σ9 incoherent twin boundaries (ITB) has been observed in TEM experiments[22,33].

The growth of a grain containing two differently oriented coherent twin boundaries (CTBs) often creates twin intersections (twin junctions), which are usually accompanied by the formation and growth of an additional GB to accommodate the misorientations across the two CTBs[34]. Twin boundaries are Σ3 GBs (where Σ represents the reciprocal coincident site lattice density related to the misorientation of the two grains/domains) and the Σ-value of the GB at the vertex is a multiple of three, depending on how many CTBs meet at the twin junction[35,36]. Because of the fixed geometry of CTBs, the set of twin-junctions which may form is limited. Orientation mapping following a grain growth annealing on Cu[37] demonstrate that the most common twin-junctions are the Σ3-Σ3-Σ9 δ-type, where the CTBs meet with dihedral angle 107°, and the Σ3-Σ3-Σ9 β-type, where CTBs meet with 70° (the designation of junction types follows Ref.[34]). Because all three grains which meet to form these junctions share a common [110] zone axis, all their atomic columns may be resolved simultaneously during TEM observation, making them ideal systems for studying TJ migration mechanisms during *in situ* experiments. Growth of CTBs by migration of incoherent twin boundaries has been experimentally observed[22,33], but migration of CTBs and Σ9 GBs as members of these common twin-junctions remains uninvestigated.

Here, we perform state-of-the-art TEM *in situ* heating experimental techniques to perform a time-resolved analysis of the coupled migration of the GBs comprising β- and δ-type twin-junctions. δ-type junctions were observed to be mobile while β-type TJs are sessile. Motion of δ-type twin-junctions (and eventual detwinning) occurs through strain-free migration of one CTB by multiples of three (111) layers towards a β-type twin-junction. The ultrafast disconnection nucleation migration mechanisms, which were not observed directly *in situ*, were studied using molecular dynamics (MD) simulations. Two migration mechanisms for δ-type twin-junctions are proposed based on disconnection theory, depending on the driving force: (1) strain-free migration via three-layer thick pure-steps nucleated from TJs under a chemical potential jump or (2) strain-accumulating migration via nucleation of partial dislocations from TJs or at the free edges of the TEM sample under a shear stress. MD simulations of Cu verify both mechanisms. Theory and MD results are used to infer that a pure step migration mechanism is responsible for the migration observed in our experiments, where migration occurs via a three (111) high step leading to no strain accumulation. This is verified by examination to twin-junction networks near the sample edge.

***In situ* observation of δ-type twin-junction migration**

In order to observe the dynamics of triple junction migration, a free-standing, ~10 nm thick, <110> fiber-textured Cu film was heated to 300°C (~0.4 melting point) *in-situ* in a high resolution TEM (HRTEM) - see Methods for details. The evolving structure was recorded at 10 frames per second (100 ms temporal resolution). Regions of the sample containing the TJs of interest were selected for observation and characterization. HRTEM images in Fig. 1 show one such region containing both a β-type and a δ-type twin-junction joined by a common Σ9 GB. All grains shared a common <110> zone axis[34], allowing all GBs/TJs to be simultaneously resolved at atomic resolution. During thermal annealing, the GB- and TJ-migrate to decrease the system energy[5]. The δ-type junction (and T3 twin boundary, labeled in Fig. 1) moves by nine {111} atomic layers towards the β-type junction between frames Fig. 1a and 1b. This change reduces the area of the Σ9 GB, the highest energy GB in this system. The junctions, separated by six {111} atomic layers did not move for the 100.8 s between the frames in Figs. 1b and 1c but migrated together and annihilated within the following ~0.1s (Fig. 1d) leaving behind the T4 CTB.

Because of the limited temporal resolution of the experiment relative to disconnection velocities during annealing, each image represents a single static period during the system evolution; i.e., the atomistic mechanism of TJ migration was not directly observed.

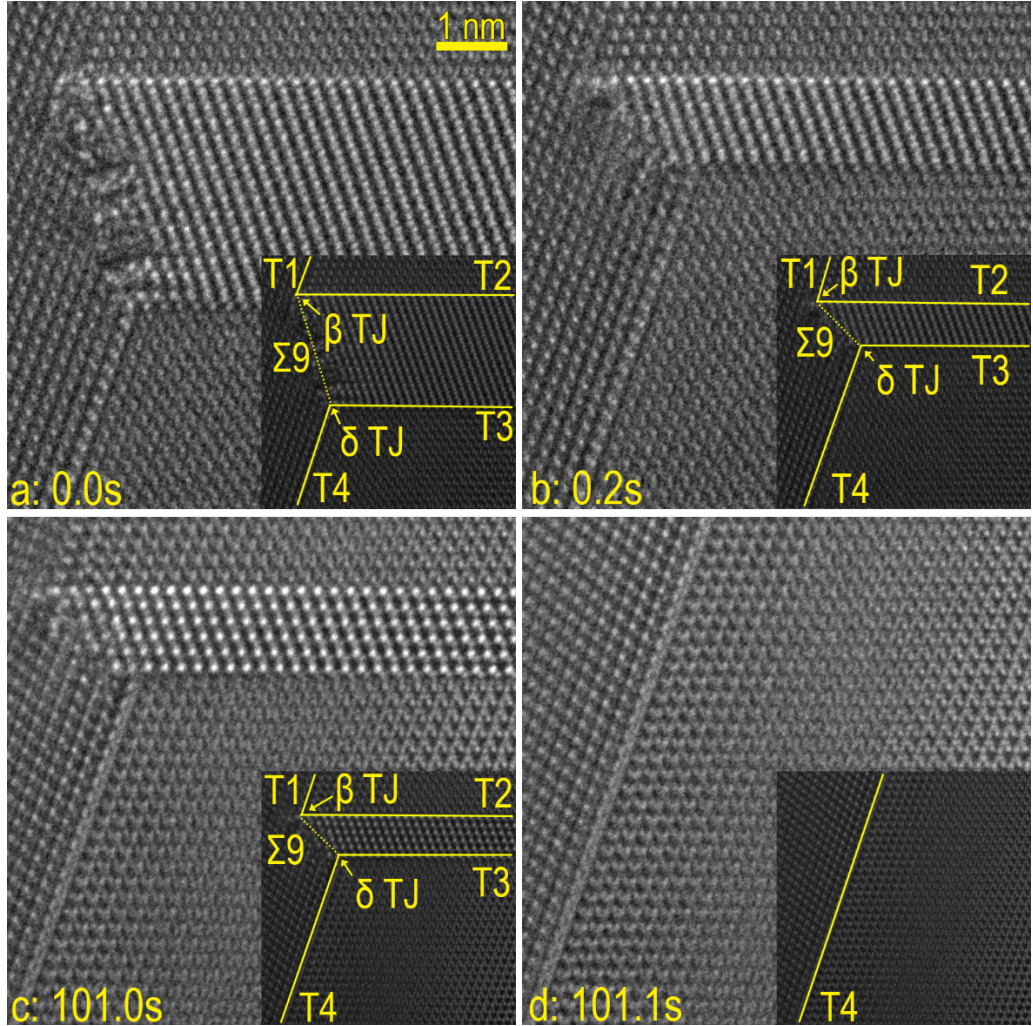

**Figure 1:** Several frames from the *in situ* HRTEM heating experiment videos showing the motion of two triple junctions. The insets show the Σ9 GB, four CTBs, and two twin-junctions. **a** The twinned region between T2 and T3 is initially 15 {111} atomic-layers thick. **b** At *t* = 0.2 s, the migration of the δ-type twin-junction (and T3 CTB) reduces the thickness of the twinned region from 15 to 6 layers and rotates the Σ9 GB. **c** The GB network remains stable for the next 100.8 s. **d** Further δ-type junction migration eliminates the twinned region leaving behind a single T4 CTB.

**Molecular dynamics simulation of TJ motion**

Given the limited temporal resolution of the *in-situ* HRTEM observations, molecular dynamics (MD) simulations were performed to clarify the observed migration mechanisms. Cylindrical simulation cells were constructred to represent the TJ and CTB geometries observed in the experiments, as shown in Fig. 2a (see Methods for details). Three configurations were studied: (i) a single CTB (Fig. 2b), (ii) a δ-type TJ (Fig. 2c), and (iii) a β-type TJ (Fig. 2d). The coordinate system was chosen such that the *x*-axis is parallel to the GB tilt axis and the *y*-axis is parallel to the Σ9 GB in Figs. 2c and d. The twin boundary in Fig. 2b is identical to the horizontal boundary in Fig. 2d. All simulations were performed at 1000 K (~0.74 melting point; higher than the experiments to accelerate the kinetics). In the simulations of stress-driven GB migration (see below), a constant shear strain rate $\dot{\varepsilon}_{xy} = 5 \times 10^9$ /s was applied parallel to the Σ9 GB plane in the tricrystal configurations (Figs. 2c and d) or a plane inclined with respect to the twin boundary in the bicrystal configuration (Fig. 2b). Additional MD simulations were performed where B migration was driven by a chemical potential jump $\psi$. See Methods for more details.

Before analyzing triple junction migration, we first focus on the migration of a solitary twin boundary. The vicinal twin (slightly off the perfect CTB inclination) of Fig. 2b was constructed and then relaxed at $T = 1000$ K for 1 ns. After relaxation, the vicinal twin relaxes into large regions of sharp coherent twin boundaries interrupted by pure steps (disconnections with zero Burgers vector) of step height corresponding to three-{111} atom layers, as seen in Fig. 2b. These pure steps migrate to the edges of the cylindrical simulation cell under a chemical potential jump, but do not respond to an external stress since they have zero net Burgers vector. This observation challenges the notion that shear stress can drive pure steps composed of 3 different partials[3].

Figure 3a-d shows that the Σ3-Σ3-Σ9 δ-type TJ migrates by the motion of three-layer-thick pure steps along one of the CTBs under a chemical potential jump (also see Fig. 2c). These pure steps nucleate at the TJ and then migrate along the horizontal CTB to the sample edge. Since pure steps have zero Burgers vector, their nucleation and migration yield zero net macroscopic strain, consistent with our experimental observations and previous experiments in Ag[33]. The diagonal (lower left) CTB in Fig. 3a-d does not move since there is no chemical potential jump across it.

The TJ motion is controlled by the motion of the horizontal CTB (and therefore, the nucleation and migration of pure steps); the mobility of the CTB is much lower than that of the Σ9 GB.

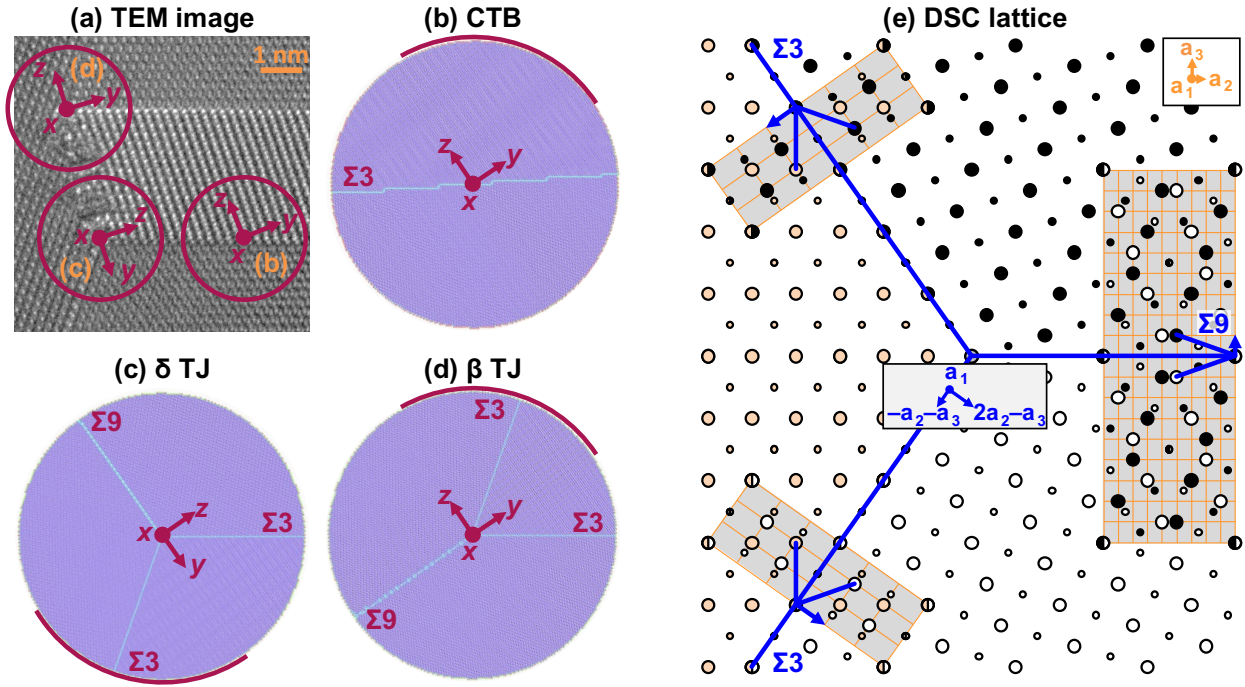

**Figure 2: a** TEM image of GB network containing δ- and β-type TJs based upon which the MD simulation cells were constructed. The comparable areas of the three simulation cells are circled, with the orientation of the simulation cell coordinate system shown. The orange labels refer to the simulation geometries in **b-d**. **b** Twin bicrystal in MD simulations, constructed with an inclination of 4° relative to the coherent twin boundary inclination. **c, d** δ- and β-type Σ3-Σ3-Σ9 TJs in MD simulations. The Σ9 GB lies in the *x-y* plane. Atoms are colored by the centrosymmetry parameter[38]; green lines are GBs. The red arcs represent fixed regions of the circular surface. **e** Lattice structure of a tricrystal with a δ-type TJ. Coincidence site and DSC lattices are shown by the orange lines in the gray rectangles for each GB (each gray rectangle shows two CSL unit cells). The DSC lattice vectors for the Σ9 GB $a_1$, $a_2$ and $a_3$ are defined.

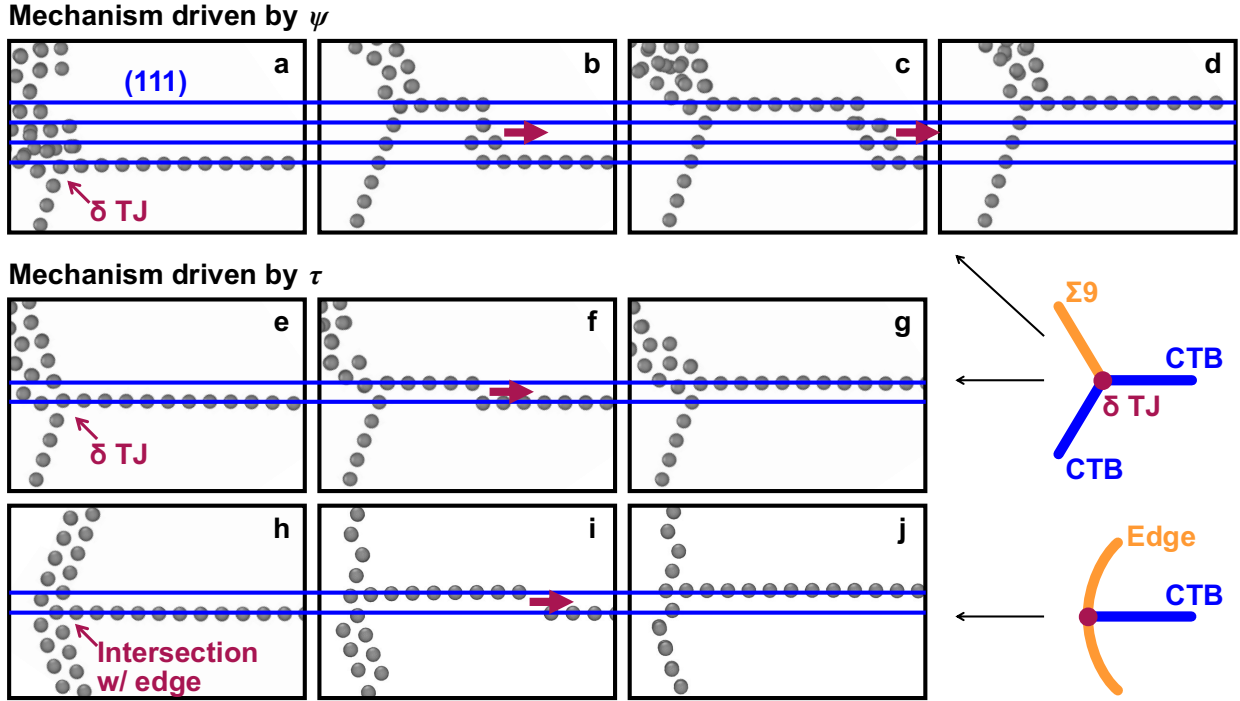

**Figure 3** Atomic-scale mechanism of δ-type TJ motion. Only atoms with centrosymmetry[38] parameter greater than 2 are shown for clarity. The red arrows indicate the direction of the disconnection motion on the horizontal CTB. **a-d** One step of δ-type TJ motion under chemical potential jump. **a** Initial position of TJ. **b, c** Nucleation and migration of 3-layer thick pure-step disconnection at the TJ. **d** The TJ and horizontal twin move upwards by three (111) layers. Migration mechanisms of δ-type TJ (**e-g**) and the intersection between a CTB and edge (**h-j**) under shear stress. **e, h** Initial position of the TJ/intersection with edge. **f, i** Nucleation and migration of 1-layer thick partials on the CTB. **g, j** The CTBs move upwards by a single (111) layer.

Figure 3e-g shows the migration of the Σ3-Σ3-Σ9 δ-type TJ under an applied stress; the TJ moves because of the horizontal CTB migration. The CTB moves, in turn, because of the motion of disconnections on it. In this case, the disconnections are not pure steps, but have both a finite step height (equal to a single {111}-atomic layer) and finite Burgers vector $(a_0/6)$**<112>** (corresponding to an FCC partial dislocation). The applied shear creates a Peach-Koehler force on the disconnection Burgers vector; $\tau$ is the shear stress resolved on to this CTB (the applied shear strain rate $\dot{\varepsilon}_{xy}$ in the coordinate system of Fig. 2c). Figure 3h-j shows the junction between a horizontal CTB and the cylindrical sample edge. These two examples show that disconnections may form at either the δ-TJs (Fig. 3e) or CTB/edge junctions (Fig. 3h) and then migrate along the CTB boundary under a shear stress driving force. This mechanism leads to macroscopic plastic strain. On the other hand, the β-type TJ in Fig. 2d does not move under either type of driving force in our MD simulations. This is consistent with our *in-situ* HRTEM experimental observations.

**Disconnection model for Σ3-Σ3-Σ9 δ-type triple junction**

Grain boundary motion occurs through the glide of disconnections; line defects with step and dislocation character[5]. In the CTB case, these disconnections can be pure steps (zero Burgers vector) or a combination of steps and Burgers vectors (in this case, the Burgers vectors are FCC Shockley partials with a Burger vector $(a_0/6)<112>$). In earlier work[5], we demonstrated that, for any GB (with fixed bicrystallography), multiple types/modes of disconnections are geometrically allowed; a mode $m$ disconnection is described by $(\mathbf{b}_m, h_m)$. For the δ-type TJ, the Σ9 displacement-shift-complete[39] (DSC) lattice basis vectors are $(\mathbf{a}_1, \mathbf{a}_2, \mathbf{a}_3)$ – see Fig. 2e. For a Σ9 GB, the three independent disconnection modes for the Σ9 GB are

$$\begin{aligned}
\mathbf{b}_1^{\Sigma 9} &= \mathbf{a}_1 & \mathbf{h}_1^{\Sigma 9} &= 4.5 N_1 \mathbf{a}_3 \\
\mathbf{b}_2^{\Sigma 9} &= \mathbf{a}_2 & \mathbf{h}_2^{\Sigma 9} &= (-1 + 4.5 N_2)\mathbf{a}_3, \\
\mathbf{b}_3^{\Sigma 9} &= \tfrac{1}{2}(\mathbf{a}_1 + \mathbf{a}_2 + \mathbf{a}_3) & \mathbf{h}_3^{\Sigma 9} &= (2 + 4.5 N_3)\mathbf{a}_3
\end{aligned} \quad (1)$$

where $N_i$ is any integer. The three independent modes for the CTB are

$$\begin{aligned}
\mathbf{b}_1^{\Sigma 3} &= -\mathbf{a}_2 - \mathbf{a}_3 & \mathbf{h}_1^{\Sigma 3} &= (1 + 3N_1)(2\mathbf{a}_2 - \mathbf{a}_3) \\
\mathbf{b}_2^{\Sigma 3} &= \tfrac{1}{2}(\mathbf{a}_1 + \mathbf{a}_2 + \mathbf{a}_3) & \mathbf{h}_2^{\Sigma 3} &= (1 + 3N_2)(2\mathbf{a}_2 - \mathbf{a}_3). \\
\mathbf{b}_3^{\Sigma 3} &= 2\mathbf{a}_2 - \mathbf{a}_3 & \mathbf{h}_3^{\Sigma 3} &= (2 + 3N_3)(2\mathbf{a}_2 - \mathbf{a}_3)
\end{aligned} \quad (2)$$

All other disconnection modes for Σ9 GBs and the CTBs are linear combinations of these three (on each). The special cases of $\mathbf{b}_1^{\Sigma 3}$ with $N_1 = 0$, $\mathbf{b}_2^{\Sigma 3}$ with $N_2 = 0$, and $-\mathbf{b}_1^{\Sigma 3} - \mathbf{b}_2^{\Sigma 3}$ with $N_1 = -1$ and $N_2 = 0$ correspond to the three classical "twinning partials".

For the TJ to migrate, the Σ9 GB and CTB must be mobile. Therefore, we ignore $(\mathbf{b}_3, \mathbf{h}_3)$ in Eqs. (1) and (2) since they are sessile (require diffusion to move). The TJ can only move if it does not accumulate Burgers vector as part of the migration process[7]. This requires the activated Burgers vector on the CTB be compatible with those on the Σ9 GB. Compatibility is satisfied only when

$$N\mathbf{b}_1^{\Sigma 9} = N\left(\mathbf{b}_1^{\Sigma 3} + 2\mathbf{b}_2^{\Sigma 3}\right), \quad (3)$$

where $N$ is integer. $N = 0$ corresponds to the case of pure steps (characterized by zero Burgers vector) for which Eq. (3) is automatically satisfied. However, this is inconsistent with stress-driven CTB migration or partial dislocation related migration. $N \neq 0$ corresponds to disconnections with Burger vector parallel to the TJ line; $\mathbf{b}_1^{\Sigma 3}$ and two $\mathbf{b}_2^{\Sigma 3}$ are emitted sequentially along the CTB. So, the TJ can continuously move when pure steps or disconnections with $\mathbf{b}\|\mathbf{a}_1$ are activated along the CTB <u>and</u> the Σ9 GB.

The most probable disconnection mode corresponds to that with the lowest energy barrier. Following our earlier work[4–6,8–10], the barrier for a ($\mathbf{b}$, $h$) mode is roughly

$$E^* = 2\left(Kb^2 + \gamma|h| + C\right)d - \left(\psi h + \tau b\right)Rd/2, \quad (4)$$

where $K = -\mu[(1 - \nu \cos 2\alpha)/4\pi(1 - \nu)] \ln[\sin(\pi r_0/R)]$, $\mu$ is the shear modulus, $\nu$ is the Poisson's ratio, $\alpha$ is the angle between the Burgers vector and the disconnection line direction, $r_0$ is the disconnection core size, $R$ is the radius of the cylindrical tricrystals, $d$ is thickness, $\psi$ is chemical potential jump, $\tau$ is shear stress, $\gamma$ is the step energy, and $C$ represents the disconnection migration barrier (associated with atomic-scale friction). The first term in Eq. (4) suggests that we should focus on disconnection modes with small $b$ and small $h$. The smallest pure step height on the CTB is $3|2\mathbf{a}_2\text{-}\mathbf{a}_3|$ ($N = 0$ in Eq. (3)), corresponding to three {111}-atomic layer thickness steps; the motion of such disconnections generates no macroscopic strain, consistent with the MD simulation driven by an applied chemical potential jump. The smallest step height of any disconnection on a

CTB is $\mathbf{b}_1^{\Sigma 3}$ (or $\mathbf{b}_2^{\Sigma 3}$) is $|2\mathbf{a}_2-\mathbf{a}_3|$ ($N = 1$ in Eq. (3)), i.e., a single {111}-atomic layer thickness, consistent with the MD result with a shear stress applied. The second term in Eq. (4) is the work done by the driving forces. The applied driving forces tilt the barrier $E^*$ and, thus, alter the most probable disconnection mode. For a disconnection mode with large $h$ and small $b$, such as a pure step, the energy barrier is lowered by application of a chemical potential jump $\psi$. For the mode with small $h$ and large $b$, such as the twinning partial mode, the barrier can be lowered by application of a shear stress. This analysis explains why the pure step mechanism is favored when CTB motion is driven by a chemical potential jump (Fig. 3a-d). In contrast, the partial dislocation Burgers vector mechanism is favored when the CTB is driven by a shear stress (Fig. 3e-g) as seen in our MD results and also observed experimentally in Cu and Ag[33]. Most deformation twins in nanocrystalline metals yield zero net macroscopic strain[33] because the chemical potential jump from large curvature is the major driving force.

**Discussion**

*In situ* TEM results were collected with a temporal resolution of 100 ms, which is slow relative to the velocities of nucleated disconnections during annealing, as discussed above. This makes it nearly impossible to directly image the disconnections responsible for evolution of the GB network. Therefore, careful analysis of the experimental results, in light of the mechanisms suggested through MD and theory, is necessary to determine the mechanism of triple junction migration observed in the experiments. There are two key distinctions between the two available migration mechanisms of the δ-type triple junction. The pure-step mechanism is strain free (zero Burgers vector) and must occur by steps with heights of multiples of three (111) layers; in contrast, the twinning-partial mechanism results in plastic deformation due to its finite Burgers vector and may occur with steps that are not multiples of three (111) layers. Examination of the steps that form on the edge of the sample during the egress/ingress of disconnections from the sample (Fig. 4) provides a means of distinguishing between the competing disconnection mechanisms.

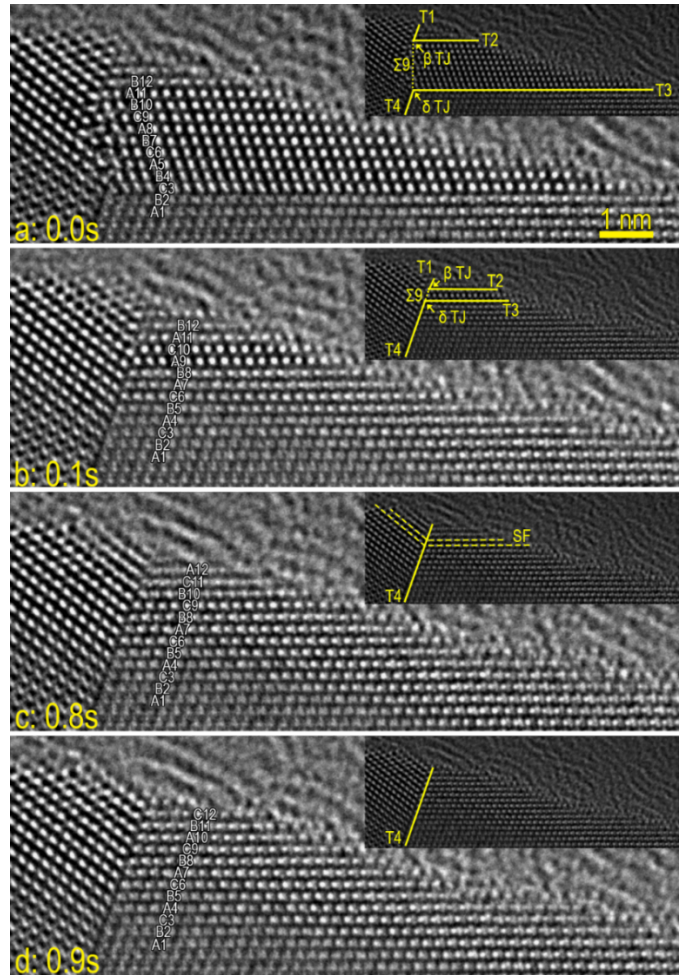

**Figure 4:** Key frames from *in situ* heating experiment videos showing motion of two triple junctions near the edge of the film. Note the change in contrast as the film edge is approached; this is indicative of a thickness gradient. A selection of (111) planes have been numbered and labeled with the FCC stacking sequence, which reverses at twin planes. The insets emphasize the boundaries and defects present. **a** The initial configuration of the local GB network, showing the β- and δ-type junctions connected by a shared Σ9 GB. **b** In the subsequent frame at $t = 0.1$ s, the δ-type junction has moved upwards by six (111) layers, reducing the length of the Σ9 GB. **c** At $t = 0.7$ s, the junctions merge; a pair of stacking faults which cross T4 remain. **d** The final state of the GB network, with only T4 persisting.

The twinned region between T2 and T3 was initially eight layers thick in the experimental images in Fig. 4. The δ-type twin-junction moves by six layers – a multiple of three – towards the β-type twin-junction, leaving behind a twinned region two layers thick. The stacking faults observed in Fig. 4c are an expected consequence because the initial thickness of the twin between two parallel CTBs is not a multiple of the step height of a pure step. The observation shown in Fig. 1 also conforms to expectations from the pure-step mechanism – the δ-type twin-junction moved by nine layers in Fig. 1b, and then by six layers in Fig. 2d; in this instance, no stacking fault remains since the thickness propagated by the CTB is a multiple of three layers. All of our observations of δ-type triple junction migration are consistent with pure step motion (i.e., CTB shifts by multiples of three (111) layers). The region shown in Fig. 4 was near the edge of the film, allowing for observation of slip steps at the edge of the film. By comparing the shape of the edge of the film after each change in the GB network, no slip steps/macroscopic strain form as the result of triple junction migration. This further reinforces the conclusion that the migration of δ-type TJ occurred by a pure-step mechanism.

We note that the GB configuration in Fig. 4b was stable for a much shorter time than the similar configuration in Fig. 1b (0.7 s versus 100.8 s, respectively). The reduced duration may be understood by consideration to terms of geometric factors. Figure 4 shows that the film is thinner near its edge than its interior (as indicated by thickness fringes) and that the T3 CTB does not meet the sample edge at a right angle. Therefore, the upward migration of this CTB (in Fig. 4) reduces the CTB area by moving it into a thinner region and shortens the CTB in the observed plane of the sample. This T3 migration also reduces the area of the high-energy Σ9 GB. Both of these effects tend to destabilize the flat CTB near a free surface (as observed in Fig. 4). The edge of a thin film is not the only possible configuration where such thickness gradients might drive TJ motion. The presence of a free surface on a sample affects GB motion in sample geometries other than the thin film geometry shown here. For example, an emerging GB meeting a surface in a bulk material will tend to migrate or rotate to intersect at 90˚. This can create curvature in an otherwise flat GB and lead to curvature driven migration. Such effects are common and well known – especially for polycrystalline particles.

The results of this study confirm that GB migration mechanism can be fundamentally different, depending on the nature of the driving force.[5] In the present CTB case, we demonstrated that changes in driving force can lead to shear coupled migration that produces macroscopic deformation/responds to stress and migration that occurs through the motion of pure steps which produce no macroscopic deformation and is not stress-driven. Previous TEM observations of disconnection-mediated GB network evolution have used mechanical loading to instigate GB migration; this necessarily implies different GB migration mechanisms than that elucidated in our experiments.

In conclusion, disconnection theory suggests that multiple modes/local mechanisms for GB migration are possible and that the mode selection is affected by the nature of the driving force for migration.[5] While the present study (*in situ* HRTEM observations and MD simulations) shows CTB migration through the motion of pure steps (no applied stress), other experimental observations (and our simulations) show migration through the motion of disconnections with a non-zero Burgers vector; these are pure-step and twinning-partial CTB migration mechanisms. The present *in situ* HRTEM observations and MD simulations demonstrate how GB migration is modified by the motion of other GBs where they meet at triple junctions (TJs); i.e., such junctions are essential features of (nearly) every polycrystalline and nanocrystalline material. The motion of GBs can "drag" their delimiting TJs and force the motion of one or more of the other GBs meeting at the TJ. In the present case, the junctions where the CTBs meet at 107° (δ-type, the third TJ member is a Σ9 GB) readily migrate while β-type TJs (CTBs meeting at 70°) are immobile. This observation is identical in both experiments and simulations and are consistent with our disconnection theory-based analysis that considers the balance of disconnection fluxes at the TJs. The easy motion of the δ-type TJ is also consistent with the relatively high mobility of the Σ9 GB and the lack of migration of the Σ9 G in the β-type TJ case, which demonstrates that unless there is disconnection flux balance at the TJ even mobile boundaries will not move. Finally, our CTB results demonstrate that disconnections may form at GB TJs and at junctions with free surfaces.

**Methods**

**Sample preparation**

Thin films <110> textured Cu films were grown by physical vapor deposition on <111>-oriented NaCl substrates to control growth direction. The films and substrate were attached, film side down, to Protochips Heating E-chips with M-BOND™ 610 epoxy-phenolic adhesive, and cured at 100°C for 3 hours. The chips, the bonded film and substrate were placed in distilled water for 5 minutes to dissolve the substrate, then the chips with attached films were placed in a fresh distilled water bath to dilute any remaining salt solution. The remaining water was evaporated under vacuum.

***In situ* heating experiments**

The heating chip was loaded into a Protochips Fusion double-tilt heating holder. HRTEM images were obtained using a spherical aberration-corrected TEM. After locating grain boundary networks suitable for observation, the sample was tilted such that all grains shared a common <110> zone axes parallel to the electron beam – to allow simultaneous resolution of all atomic positions and edge-on grain boundary/triple junction structures. The sample was then heated from room temperature to 300°C, at a heating rate of 1000°C/s and held at that temperature for the duration of the experiment. Grain boundary/triple junction migration were recorded using the Gatan OneView charge-coupled device (CCD) camera at a temporal resolution of 100 ms. All experiments were performed at 300 kV.

**Molecular dynamics simulations**

MD simulations were performed using the Large-scale Atomic/Molecular Massively Parallel Simulator (LAMMPS)[40] and a Cu embedded-atom-method potential[41]. Cylindrical tricrystals were constructed with radius $R \sim 50$ nm and thickness $d \sim 1$ nm. Periodic boundary conditions were applied in the direction perpendicular to the plane of the cylinder. A cylindrical twin boundary bicrystal was constructed with an inclination angle of 4° with the same simulation size. Before applying a driving force, the system was equilibrated at 1000 K for 1 ns. In the simulations of stress-driven GB migration, a constant strain rate $\dot{\varepsilon}_{xy} = 5 \times 10^9$ /s was applied. The circular sample surface was free except for a 90° fixed arc, indicated in red in Fig. 2b-d. Simulations of chemical potential difference-driven GB migration were performed with a synthetic driving force[42] of $\psi = 0.01$ eV/atom. The synthetic driving force was added to the atoms in the grain opposite to the fixed arc. All simulations were run for 2 ns at 1000 K at a fixed number of atoms and temperature (Nosé-Hoover thermostat on all but the fixed atoms). Atoms are colored by the centrosymmetry parameter[38] in Fig 2. Supplementary Fig. 1 shows the construction of the δ-type twin-junction in Fig. 2c.

## Acknowledgements

This work was supported by the Army Research Office (ARO) under Grant W911NF-19-1-0263. The authors acknowledge the use of facilities and instrumentation at the UC Irvine Materials Research Institute (IMRI) supported in part by the National Science Foundation through the Materials Research Science and Engineering Center program (DMR-2011967).

## Data availability

The data supporting the findings in this study are available upon reasonable request from the corresponding author.